\documentclass[lettersize,journal]{IEEEtran}
\usepackage{amsmath,amsfonts}
\usepackage{algorithmic}
\usepackage{algorithm}
\usepackage{array}
\usepackage[caption=false,font=normalsize,labelfont=sf,textfont=sf]{subfig}
\usepackage{textcomp}
\usepackage{stfloats}
\usepackage{url}
\usepackage{verbatim}
\usepackage{graphicx}
\usepackage{cite}
\title{A Kalman Filter-Assisted Data-Predictive SAR ADC With Reduced Switching Energy for Low-Power Applications}

\author{
    Xiyuan~Feng\textsuperscript{1},
    Yuxiang~Zhao\textsuperscript{1},
    Jie~Xiong\textsuperscript{1},
    Dian~Lin\textsuperscript{1},
    Yunlei~Zhong\textsuperscript{2},
    Wei~Liu\textsuperscript{1},
    Zhongheng~Ji\textsuperscript{1},
    Ruiyu~Tian\textsuperscript{1},
    Chenhao~Zhuo\textsuperscript{1,*},
    and~Yue~Yin\textsuperscript{1,*}
    
    \and [0.5em] \small \textsuperscript{1}School of Integrated Circuits (School of Microelectronics), Northwestern Polytechnical University, 
    No. 1 Dongxiang Road, Chang’an District, Xi’an 710129, P. R. China.
    
    \and [0.5em] \small \textsuperscript{2}Key Laboratory of Multifunctional Nanomaterials and Smart Systems Division of Advanced Materials,Suzhou
Institute of Nano-Tech and Nano-Bionics, Chinese Academy of Sciences, 
     Suzhou, 215128, , P. R. China
    \and [0.3em] \small *Corresponding authors: Yue Yin (yinyue@nwpu.edu.cn) and Chenhao Zhuo (zhuochenhao@outlook.com)
    
    \thanks{This paper was produced by the IEEE Publication Technology Group. They are in Piscataway, NJ.}
    \thanks{Manuscript received April 19, 2021; revised August 16, 2021.}
    \thanks{This work was also supported in part by the Fundamental Research Funds for the Central Universities of Ministry of Education of China (No. D5000240188) and a fellowship from the China Postdoctoral Science Foundation (No. 2025M784413).}
}

\begin{document}
\markboth{Journal of \LaTeX\ Class Files,~Vol.~14, No.~8, August~2021}%
{Shell \MakeLowercase{\textit{et al.}}: A Sample Article Using IEEEtran.cls for IEEE Journals}

\maketitle

%\IEEEpubid{0000--0000/00\$00.00~\copyright~2026 IEEE}

\begin{abstract}
The proliferation of Internet of Things (IoT) devices and wearable health monitors has created an urgent demand for ultra-low-power analog-to-digital converters (ADCs). Successive approximation register (SAR) ADCs are widely used in such applications, yet their energy efficiency remains constrained by the sequential bit-by-bit switching of the capacitive DAC (CDAC). The high-weight most significant bit (MSB) transitions dominate the total switching energy, and the rigid $N$-cycle conversion flow imposes a hard lower bound on latency per sample.This paper presents a Kalman filter-assisted data-predictive SAR ADC that replaces the first four comparator-driven decisions with a recursive state estimator. The Kalman filter predicts the 4 MSBs from the complete conversion history before each cycle begins, enabling simultaneous parallel switching of the MSB capacitors. This eliminates redundant CDAC transitions, shortens the quantization cycle by four clock periods, and reduces switching energy by approximately 50\%. An optimized 4-bit MSB switching scheme further suppresses residual switching at the hardware level. The ADC, designed in a 180-nm CMOS process, supports configurable dual-mode operation, toggling between a conventional mode and the Kalman-driven predictive mode for robustness under erratic inputs. At 20 MS/s and a 1.8-V supply, the predictive mode reduces total power consumption by 50.3\% (from 1.96 mW to 0.975 mW), with a measured SNR/SFDR of 57.88/74.51 dB at 504 kHz, confirming its suitability for energy-constrained wireless sensor networks.
\end{abstract}


% ============================================================
% SECTION I: Introduction
% ============================================================

% Section I: Introduction
\section{Introduction}

ADCs are essential building blocks that bridge the physical analog world and the digital processing domain. With the rapid proliferation of the IoT, wearable health monitors, and energy-harvesting sensor nodes, there is an unprecedented demand for ultra-low-power data conversion. In these battery-constrained systems, the ADC often dominates the total hardware power budget; maximizing its energy efficiency is therefore critical to extending operational lifetime~\cite{1,2}.
\IEEEpubidadjcol

Among various ADC topologies,  SAR ADCs are widely adopted for energy-constrained applications due to their predominantly digital architecture, absence of power-hungry operational amplifiers, and excellent inherent hardware efficiency. Nevertheless, the conventional $N$-bit SAR ADC is fundamentally bottlenecked by its sequential bit-by-bit trial-and-error procedure. The capacitive digital-to-analog converter (CDAC) must undergo multiple successive switching cycles for every sample. As the resolution increases, this rigid step-by-step conversion leads to a prolonged quantization cycle and substantial switching energy, particularly during the transitions of the high-weight MSB capacitors. Consequently, the latency and dynamic power inherent in traditional CDAC switching algorithms impose a severe limit on further efficiency improvements.

To mitigate redundant CDAC transitions, numerous switching schemes have been proposed. Merge-and-split switching~\cite{1}, VCM-based switching~\cite{2}, and one-side switching instead combined with higher-bit switching instead (OSSI+HBSI)~\cite{3} optimize the input sampling strategy, reference-voltage usage, or bottom-plate switching sequence. While these methods reduce the average switching energy, they still follow the conventional binary-search flow: an $N$-bit conversion requires $N$ comparison cycles, and the CDAC residue must settle within one least significant bit (LSB). Moreover, certain schemes introduce common-mode voltage variations that can degrade linearity and cause output distortion~\cite{3}.

An alternative line of research exploits the predictability of the input signal to accelerate SAR conversion. Representative approaches include: employing additional reference comparators~\cite{4}, incorporating time-domain quantization into the SAR loop~\cite{5,6}, using a coarse ADC for signal pre-estimation~\cite{7}, introducing dedicated analog prediction structures~\cite{8}, and applying window-based detection to determine whether the input falls within a predefined amplitude range~\cite{9,10,11}. Although effective for slowly varying or biomedical signals, these methods typically incur considerable hardware overhead, increased control complexity, or degraded robustness under abrupt signal variations and high-frequency input components.

To address these limitations and break the trade-off between predictive accuracy and conversion speed, a promising direction is to predict the upcoming MSBs through purely digital-domain computation, thereby avoiding the analog hardware overhead of the methods surveyed above. Wood and Sun pioneered this digital prediction approach~\cite{18}, employing a first-order difference equation to estimate the top $K$ bits of the next sample from the two most recent conversion values. However, the first-order predictor relies on only two historical samples, and its accuracy degrades sharply under signal transients where the local linearity assumption fails. Building on this foundation, this paper replaces the first-order predictor with a recursive Kalman filter that exploits the complete conversion history, adaptively tracking both the signal amplitude and its local rate of change. The resulting Kalman filter-assisted data-predictive SAR ADC, implemented in 180-nm CMOS technology, provides robust MSB prediction even under dynamic signal conditions. The primary contributions are summarized as follows:

\begin{enumerate}
\item \textbf{Kalman Filter-Assisted MSB Prediction:} A Kalman filter is embedded into the SAR logic to dynamically predict the switching values of the first 4 MSB capacitors based on historical conversion trajectories. This predictive framework provides robust tracking even under dynamic signal conditions.

\item \textbf{Simultaneous MSB Switching:} By executing the prediction prior to comparison, the proposed design enables the simultaneous, parallel switching of the first 4 MSB capacitors. This eliminates superfluous CDAC transitions, reduces dynamic switching energy, and shortens the quantization cycle by 4 clock periods.

\item \textbf{Optimized 4-Bit Switching Logic:} An improved 4-bit MSB switching technique is introduced to further eliminate residual redundant switching events, maximizing hardware-level power savings.

\item \textbf{Configurable Dual-Mode Operation:} The architecture supports seamless toggling between a traditional mode (filter off) and a predictive mode (filter on). This dual-mode capability safeguards system robustness under unknown or highly erratic signal inputs.
\end{enumerate}

The efficacy of the proposed design is validated through extensive circuit simulations under a 1.8-V supply. At 20~MS/s, the predictive mode reduces total power consumption from 1.96~mW to 0.975~mW, achieving a 50.3\% reduction. Dynamic performance simulations demonstrate an SNR of 57.88~dB and an SFDR of 74.51~dB at 504~kHz input.

The remainder of this paper is organized as follows. Section~II reviews prior art in predictive and pre-measurement SAR ADC techniques. Section~III describes the proposed architecture, including the Kalman filter prediction algorithm and the optimized 4-bit MSB switching logic. Section~IV presents the circuit implementation of the SAR control logic and DAC controller. Section~V reports the simulation results and performance comparison. Section~VI discusses limitations and future work, and Section~VII concludes the paper.

% ============================================================
% SECTION II: Prior Art
% ============================================================
% Section II: Prior Art --- Pre-Measurement and Predictive Techniques
\section{Prior Art: Pre-Measurement and Predictive Techniques}
\label{sec:prior}

In a conventional SAR ADC, the DAC switching scheme adjusts the CDAC by adding or subtracting a binary-weighted voltage at each comparator decision, progressively driving the top-plate residue voltage toward zero. By the end of conversion, the residue must fall below one LSB. However, because the comparator responds only to the polarity of the CDAC voltage, later switching steps can partially undo the charge redistribution produced by earlier ones; charging and then discharging the CDAC array is inherently inefficient. A proactive alternative is to predict the signal amplitude range before switching begins. Once the amplitude is confirmed to lie within a preset sub-range, a targeted switching strategy can be adopted that bypasses the large capacitors and engages only the smaller ones.

One approach to window-based prediction performs time-domain quantization on the CDAC voltage. Guerber \textit{et al.}~\cite{5} integrated a time comparator and a delay unit into the critical path of the SAR loop, producing a three-level output (``00'', ``01'', ``10'') that is transmitted to the SAR logic. Functioning analogously to a 1.5-bit/stage architecture, this scheme provides redundancy that tolerates minor settling errors. When a mid-code (``01'') is generated, no switching operation is executed. As conversion proceeds, the boundaries of the timing quantization levels are updated to improve energy efficiency. By converting voltage into time duration through the time comparator, this method predicts the first two bits before the corresponding switching decisions.

The Detect-and-Skip (DAS) window switching scheme~\cite{7} employs a coarse SAR ADC to resolve the first five MSBs and supplies skip-control logic that adjusts the DAC of a fine SAR ADC for the remaining bits. As noted in~\cite{7}, switching three unit capacitors (2C and 1C) sequentially in a 3-bit SAR ADC consumes a total switching energy of $1.25\,CV^2$, whereas switching them simultaneously reduces the energy to $0.75\,CV^2$. The DAS algorithm avoids unnecessary switching of the MSB capacitors in the fine DAC and is especially effective for converting small-amplitude differential input signals.

In the First 2-bit Guess (F2G) scheme~\cite{9}, the values of the first two MSBs are predicted in advance. This architecture uses an additional dummy DAC (D-DAC) to shift the top-plate voltage of the main DAC (M-DAC) upward by $V_{cm}=V_{ref}/2$. Three comparators then operate in parallel: $Q_p$ compares the original input $V_{inn}$ with its complementary offset $V_{inps}$, while $Q_m$ and $Q_n$ compare $V_{inps}-V_{inns}$ and $V_{inp}-V_{inns}$, respectively. Both the polarity and the magnitude (greater or less than $V_{ref}/2$) of the differential signal are thus determined within a single cycle, enabling the MSB and MSB$-$1 to be resolved simultaneously.

Canal \textit{et al.}~\cite{12} proposed a time-assisted SAR ADC architecture that uses a time-to-digital converter (TDC) to implement a window switching scheme. The TDC predicts the input signal amplitude range within the first SAR cycle and determines the switching values for the three MSB capacitors. The architecture combines bit-guessing with digital error correction (DEC): a 12-bit word obtained from ten SAR cycles is processed into the final 10-bit output. Key capacitors are split to provide redundancy that tolerates TDC decision errors and capacitor mismatches, and correlated reversed switching (CRS) is adopted to compensate for residual mismatch errors.

The Wood and Sun predictor~\cite{18} operates as follows. Before each conversion, the next sample is estimated using a first-order difference equation $\hat{X}[n+1] \approx X[n] + (X[n] - X[n-1])$, which extrapolates the local linear trend from the two most recent conversion results. The first $K$ bits of this estimate are forwarded to the DAC, and the corresponding MSB capacitors are held fixed, skipping $K$ comparator cycles. The remaining $B-K$ LSBs are resolved via conventional binary search, with one additional redundant LSB comparison serving as an error check. If the prediction falls outside the correctable range, the ADC reverts to a full $B$-bit MSB-first conversion, ensuring no loss of signal information. An accuracy tracking circuit monitors the largest sample-to-sample difference and adaptively adjusts $K$ to match the current signal conditions. Because the predictor consists solely of digital registers, a subtractor, and a finite-state machine, the analog front-end remains unchanged.

In summary, the methods surveyed above aim to pre-measure or predict the MSBs and then formulate optimized switching strategies that avoid unnecessary DAC transitions, thereby saving energy and shortening the conversion cycle. These approaches share a common limitation: they rely on heuristics or window-based thresholds that use only a fraction of the available conversion history. The present work extends this line of research by replacing heuristic and window-based prediction with a Kalman filter that recursively estimates the MSB values from the complete conversion history, exploiting all past comparator decisions to refine the prediction as the conversion proceeds.

% ============================================================
% SECTION III: Proposed Architecture
% ============================================================
% Section III: Proposed SAR ADC Architecture
\section{Proposed SAR ADC Architecture}
\label{sec:architecture}

\begin{figure*}[t]
\centering
\includegraphics[width=5in]{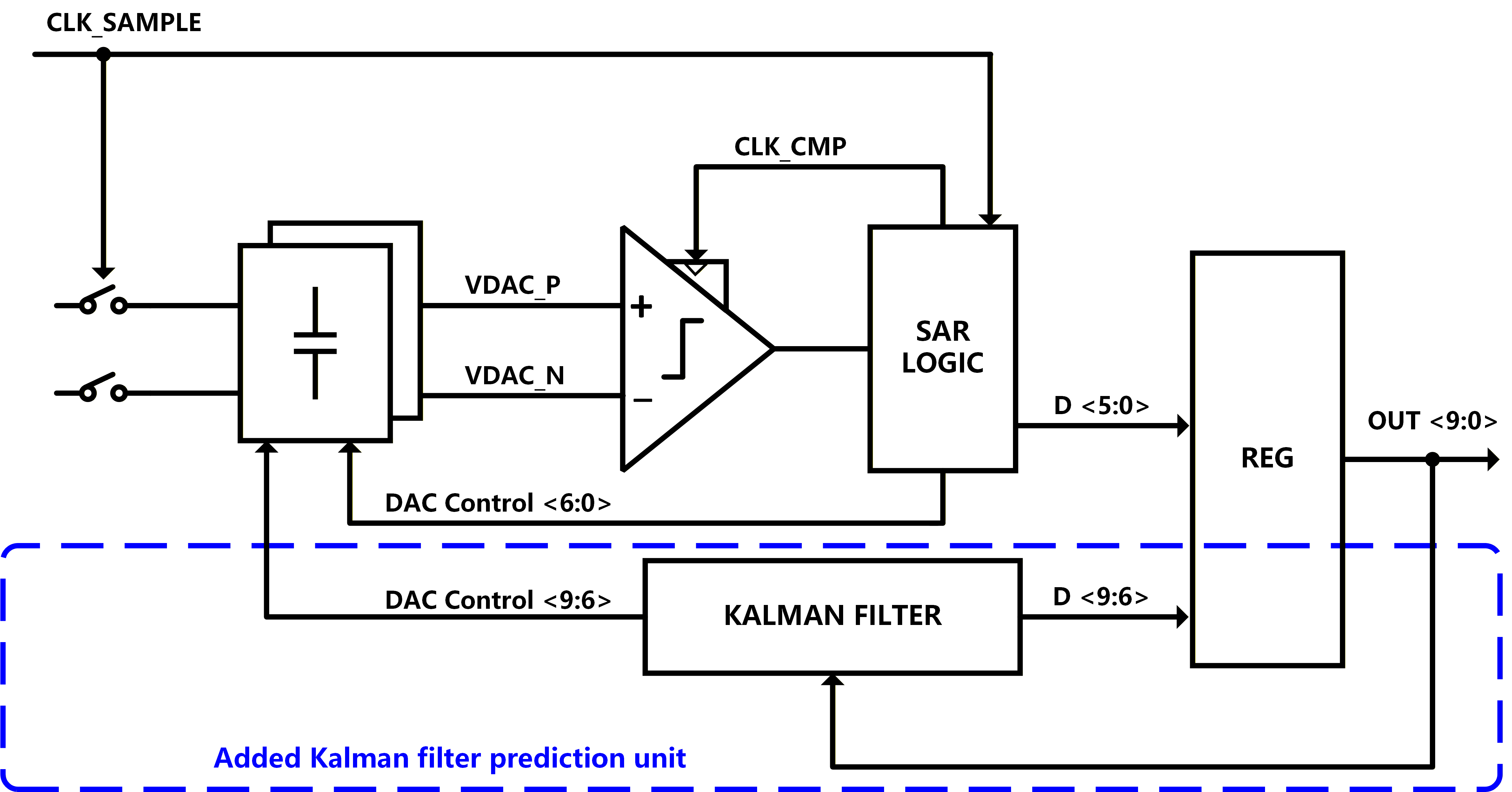}%
\caption{Block diagram of the proposed SAR ADC integrated with a Kalman filter prediction engine.}
\label{fig1}
\end{figure*}

Fig.~\ref{fig1} shows the proposed ADC architecture. The design uses differential CDAC arrays, each with a bridge capacitor to reduce the total capacitor area. The top plates of the CDAC capacitors connect to the comparator inputs. The bottom-plate switches for the $<$9:6$>$ MSB capacitors are driven by the Kalman filter output; those for the $<$5:0$>$ LSB capacitors are controlled by the conventional SAR logic.

The SAR LOGIC module generates an internal asynchronous clock from the external sampling signal CLKS and coordinates comparator operation and capacitor switching. Each conversion proceeds in three phases. In the prediction phase, the Kalman filter computes a 4-bit MSB estimate from prior conversion results, and the four MSB capacitors are switched simultaneously. In the quantization phase, the SAR logic resolves the remaining 6 LSBs via a conventional binary search. In the update phase, the completed 10-bit output is fed back to the Kalman filter to seed the next prediction cycle.

\subsection{Kalman Filter-Based Data Prediction}

The defining innovation of this work is to replace the first four comparator-driven decisions of a conventional SAR cycle with a Kalman filter that predicts those four MSBs \emph{before} conversion begins. This section explains why prediction is feasible, how the Kalman filter generates the estimate, and how the prediction is integrated into the SAR timing.

\subsubsection{Prediction Principle}

Signals encountered in biomedical and sensor applications exhibit strong sample-to-sample correlation: the value at time $k$ is rarely far from the value at $k-1$. This temporal redundancy enables a predictive approach; if the next sample can be estimated from conversion history with sufficient accuracy, the most significant bits of the estimate will already match those of the true input, and their comparison cycles can be bypassed entirely. The design problem therefore reduces to choosing how many MSBs to predict. Through MATLAB modeling across ECG and EEG waveforms, a 4-bit prediction was found to provide the optimal trade-off: it eliminates 4 of 10 comparison cycles while confining the residual $\delta(k)$ within the 6-bit correction window for over 94\% of samples.

\subsubsection{Two-State Kalman Predictor}

The predictor maintains a minimal state vector comprising the signal amplitude and its local rate of change:
\begin{equation}
\mathbf{x}(k) = \begin{bmatrix} x(k) \\ \dot{x}(k) \end{bmatrix},
\label{eq:state}
\end{equation}
where $x(k)$ is expressed in units of the 10-bit ADC output code and $\dot{x}(k)$ is in codes per sampling period. The state evolves under a constant-velocity projection, the simplest model that captures local linear trends:
\begin{align}
\hat{x}(k|k-1) &= \hat{x}(k-1|k-1) + T_s \cdot \dot{\hat{x}}(k-1|k-1), \label{eq:kf_pred_simple} \\
\dot{\hat{x}}(k|k-1) &= \dot{\hat{x}}(k-1|k-1). \label{eq:kf_pred_dot}
\end{align}
Here $T_s = 1/f_s$, and $\hat{x}(k|k-1)$ denotes the estimate of $x$ at sample $k$ conditioned on measurements up to sample $k-1$. These two equations constitute the prediction step and require only one multiply--add operation per sample. They are executed \emph{before} the $k$-th sampling edge, using only the state estimate from the previous conversion cycle; no new measurement is needed, and the prediction is ready when the sampling clock fires.

The amplitude prediction $\hat{x}(k|k-1)$ is then quantized to a 10-bit integer:
\begin{equation}
X_{kf}(k) = \operatorname{round}\!\left(\frac{\hat{x}(k|k-1)}{V_{\text{ref}}} \times 2^{10}\right).
\label{eq:quantize}
\end{equation}
The upper 4 bits of $X_{kf}(k)$, i.e., bits 9:6, are forwarded to the CDAC. The four corresponding capacitors are switched simultaneously, following the optimized scheme of Section~\ref{sec:switching}, \emph{before} the comparator is strobed for the first time. These four cycles are therefore skipped entirely.

\subsubsection{Closing the Loop: Quantization and State Update}

Once the MSB capacitors settle, the comparator senses the residual voltage $\delta(k)$ on the CDAC top plate. All voltages in what follows use the 10-bit LSB as the unit: $\text{LSB}_{10} = V_{\text{ref}}/2^{10}$. If the prediction lies within the correctable window $0 \leq \delta(k) \leq 63\;\text{LSB}_{10}$, the SAR logic resolves the remaining 6 bits via a conventional binary search. The complete 10-bit output is:
\begin{equation}
\mathit{result}(k) = \bigl\{X_{kf}(k)[9:6],\; \text{SAR\_LSB}[5:0]\bigr\}.
\label{eq:assemble}
\end{equation}

This result serves two roles simultaneously. It is the ADC's digital output delivered to the external system, and it is also the measurement $z(k) = \mathit{result}(k)$ fed back to the Kalman filter. A crucial simplification distinguishes this application from generic Kalman filtering: the ADC output \emph{is} the true digital representation of the input (to within the quantization error). There is no separate ``sensor noise'' to overcome; the ``measurement noise'' is simply the ADC's own quantization error, whose statistics are known \emph{a priori}. Consequently, the update step takes an unusually simple form:
\begin{align}
\hat{x}(k|k) &= z(k), \label{eq:kf_update_x} \\
\dot{\hat{x}}(k|k) &= \dot{\hat{x}}(k|k-1) + \alpha \cdot \bigl(z(k) - \hat{x}(k|k-1)\bigr). \label{eq:kf_update_simple}
\end{align}
Equation (\ref{eq:kf_update_x}) states that the amplitude state is directly reset to the measured ADC code; the filter trusts the measurement completely. Only the derivative state, which tracks the signal slope, accumulates filtered history through (\ref{eq:kf_update_simple}). The scalar gain $\alpha$ controls how aggressively the slope estimate responds to prediction errors. A larger $\alpha$ enables faster tracking of signal transients at the cost of increased sensitivity to quantization noise; a smaller $\alpha$ produces smoother predictions that may lag behind rapid changes. In the present design, $\alpha$ is tuned empirically via MATLAB simulation to minimize the RMS prediction error over the target signal ensemble.

Fig.~\ref{fig1} captures the resulting predict--quantize--update loop. The Kalman filter computes the prediction; the four MSBs preset the CDAC; the SAR resolves the six LSBs; the completed result $z(k)$ feeds back to correct the state estimate; and the corrected state drives the next prediction. Only two multiply--add operations per sample are on the critical path, making a dedicated hardware realization (Section~\ref{sec:discussion}) feasible with minimal digital overhead.

\subsubsection{Handling Prediction Failures}

The predictor is not infallible. When the signal undergoes an abrupt transient, or when the filter has not yet converged from initialization, the residual falls outside the 6-bit correction window. Two failure modes arise:
\begin{itemize}
\item \textbf{Under-prediction} ($\delta(k) > 63\;\text{LSB}_{10}$): the SAR saturates at ``111111''; the assembled $\mathit{result}(k)$ underestimates the true input.
\item \textbf{Over-prediction} ($\delta(k) < 0$): all comparator outputs are ``0''; the SAR returns ``000000'' and $\mathit{result}(k)$ overestimates the true input.
\end{itemize}
Both conditions yield a corrupted output and are detected by monitoring whether the 6-bit SAR terminates in an all-ones or all-zeros state. A single saturation event may arise from a noise spike and does not warrant abandoning the prediction mode. The dual-mode controller (Section~\ref{sec:control}) therefore requires \emph{two consecutive} saturation flags before triggering a fallback. Upon fallback, the ADC reverts to full 10-cycle conventional conversion for a recovery period of $M$ samples, while the Kalman filter continues to observe measurements with its prediction output gated off. After $M$ samples, the state is re-initialized from the most recent valid conversion, and prediction-driven conversion resumes.

This hysteresis-based protection is particularly effective for biomedical signals. Consider the electrocardiogram (ECG): the waveform consists of extended low-activity segments (P--R interval, S--T segment, typical duration 100--400~ms) where the signal changes slowly and the predictor achieves its highest accuracy and energy savings. These are punctuated by sharp QRS complexes (80--120~ms) that may transiently trigger the fallback. Because the recovery period $M$ is chosen to be shorter than the QRS duration, the predictor re-engages before the next low-activity segment begins, preserving the net energy benefit over the full cardiac cycle.

\begin{figure}[!t]
\centering
\includegraphics[width=3in]{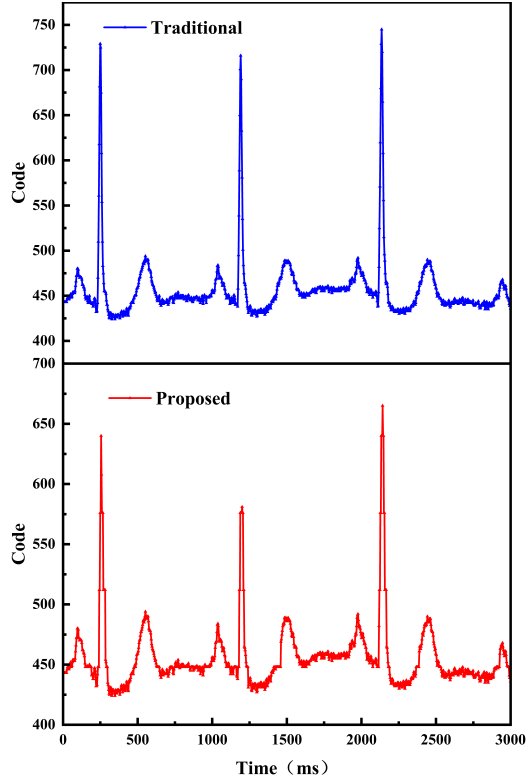}%
\caption{Simulated ADC quantization results: comparison between traditional SAR operation and the proposed Kalman-filter-assisted prediction mode for an ECG-like input waveform.}
\label{fig2}
\end{figure}

\begin{figure}[!t]
\centering
\includegraphics[width=3in]{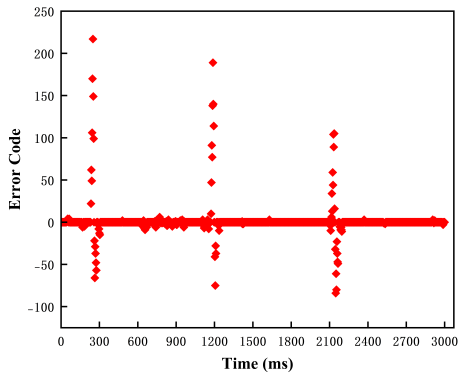}%
\caption{Quantization error before and after Kalman filter integration. The prediction mode markedly reduces error during low-frequency windows.}
\label{fig3}
\end{figure}

To illustrate the system-level prediction performance, consider ECG signal measurement as a representative application. The ECG waveform naturally decomposes into two regimes: extended low-frequency windows, where the signal varies slowly and the Kalman predictor tracks it with high fidelity, and brief high-frequency windows (QRS complexes), where large-amplitude transients challenge the prediction. The proposed architecture accurately predicts and quantizes the signal throughout the low-frequency windows. Fig.~\ref{fig2} compares the quantization output of the prediction-assisted mode against conventional SAR operation for an ECG-like input waveform; the Kalman-filter-enabled prediction closely follows the true signal during the low-activity intervals. Fig.~\ref{fig3} presents the corresponding quantization error, where the prediction mode achieves a marked reduction in error during these same intervals.

\subsection{Optimized 4-Bit MSB Switching Logic}
\label{sec:switching}

In the conventional monotonic switching strategy, the comparator decision switches one capacitor's bottom plate from $V_{\text{REF}}$ to ground at each cycle. For a 4-bit differential CDAC array, let the digital output bits be $D_3 D_2 D_1 D_0$, where $D_i \in \{0, 1\}$ and $\bar{D}_i = 1 - D_i$. After all switching steps, the differential voltage at the comparator input is given by:

\begin{equation}
\label{eq:1}
V_{\text{CDAC}} = V_{\text{IN}} - \frac{V_{\text{REF}}}{2^4}\sum_{i=0}^{3} 2^{i}\bigl(D_i - \bar{D}_i\bigr).
\end{equation}

In this process, each comparator decision triggers exactly one capacitor transition. Crucially, the switching of lower-weight LSB capacitors can partially cancel the voltage change produced by higher-weight MSB capacitors, wasting switching energy. Because the proposed architecture receives the four MSBs from the Kalman filter before the comparison phase, an optimal switching strategy can be pre-computed to achieve the target residue with the minimum number of capacitor transitions.

For example, if the target 4-bit code is ``1000'', the conventional approach sequentially flips the positive $8C$, negative $4C$, negative $2C$, and negative $1C$ capacitors, a total of four switching events. With the proposed strategy, only the positive $1C$ capacitor is flipped to establish $V_{\text{IN}} - 1\,\text{LSB}$ at the comparator input, eliminating three redundant transitions.

Table~\ref{TABLE1} enumerates the switching configurations for all 16 possible MSB codes. In the capacitor array columns, ``1'' denotes a bottom-plate switch from $V_{\text{REF}}$ to ground, and ``0'' denotes no switching. Compared with the conventional monotonic scheme, the proposed strategy reduces the total number of capacitor switching events by approximately 50\% averaged over all codes, thereby cutting the dynamic switching energy by half.

\begin{table*}[t]
\centering
\caption{Proposed optimized 4-bit MSB switching scheme versus conventional monotonic switching.}
\label{TABLE1}
\begin{tabular}{|c|c|c|c|c|c|}
\hline
Code & \multicolumn{2}{c|}{Conventional} & \multicolumn{2}{c|}{Proposed} & Target \\
     & CDACP$<3:0>$ & CDACN$<3:0>$ & CDACP$<3:0>$ & CDACN$<3:0>$ & Residue \\
\hline
0000 & 0000 & 1111 & 0000 & 1111 & $+15$  \\
0001 & 0001 & 1110 & 0000 & 1101 & $+13$  \\
0010 & 0010 & 1101 & 0000 & 1011 & $+11$  \\
0011 & 0011 & 1100 & 0000 & 1001 & $+9$  \\
0100 & 0100 & 1011 & 0000 & 0111 & $+7$  \\
0101 & 0101 & 1010 & 0000 & 0101 & $+5$  \\
0110 & 0110 & 1001 & 0000 & 0011 & $+3$  \\
0111 & 0111 & 1000 & 0000 & 0001 & $+1$  \\
1000 & 1000 & 0111 & 0001 & 0000 & $-1$  \\
1001 & 1001 & 0110 & 0011 & 0000 & $-3$  \\
1010 & 1010 & 0101 & 0101 & 0000 & $-5$  \\
1011 & 1011 & 0100 & 0111 & 0000 & $-7$  \\
1100 & 1100 & 0011 & 1001 & 0000 & $-9$  \\
1101 & 1101 & 0010 & 1011 & 0000 & $-11$  \\
1110 & 1110 & 0001 & 1101 & 0000 & $-13$  \\
1111 & 1111 & 0000 & 1111 & 0000 & $-15$  \\
\hline
\end{tabular}
\end{table*}

% ============================================================
% SECTION IV: Circuit Implementation
% ============================================================
% Section IV: Circuit Implementation
\section{Circuit Implementation}
\label{sec:implementation}

\subsection{SAR Control Logic With Prediction Interface}
\label{sec:control}

\begin{figure*}[t]
\centering
\includegraphics[width=5in]{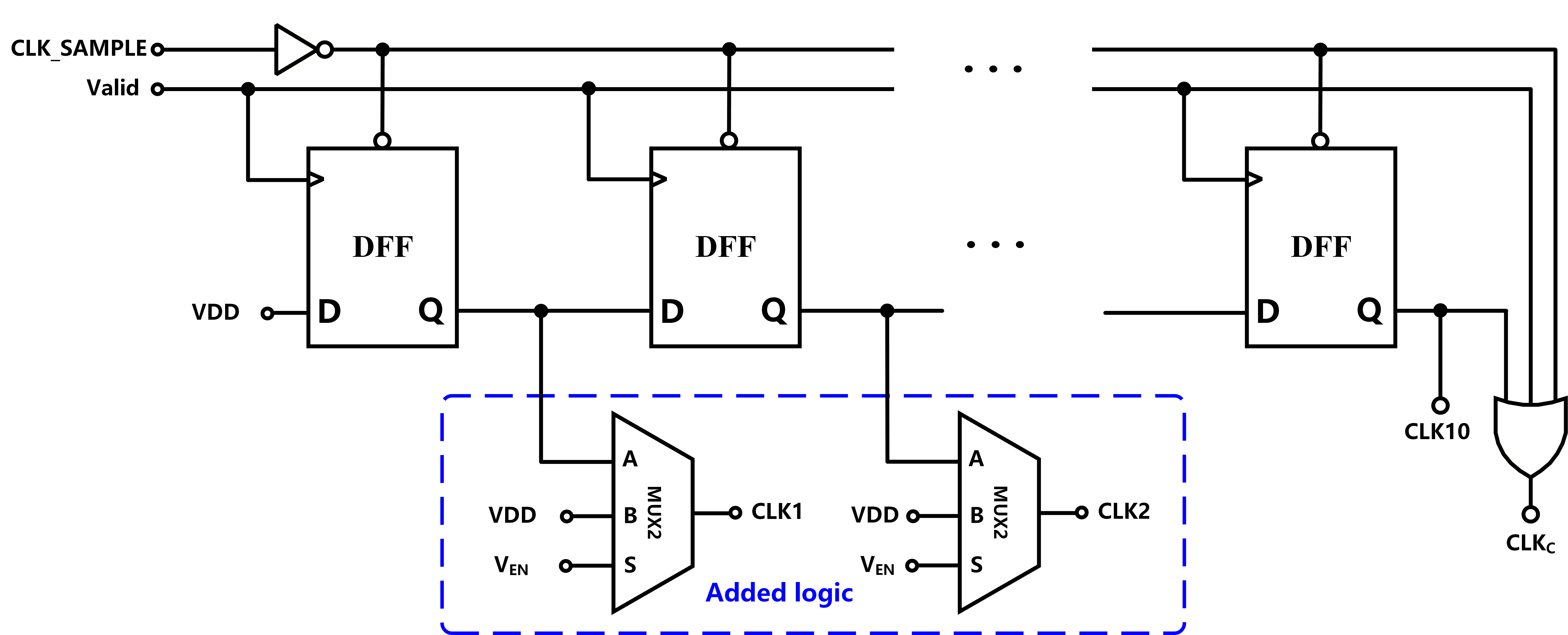}%
\caption{Proposed SAR logic with integrated Kalman filter prediction module and optimized 4-bit MSB switching controller. The MUX2 block selects between prediction-driven and comparator-driven clock paths based on $V_{EN}$.}
\label{fig5}
\end{figure*}

The SAR logic designed in this work extends a conventional asynchronous SAR controller by adding a Kalman filter data-transmission interface and dual-mode control. Fig.~\ref{fig5} shows the proposed architecture, with the prediction path operating alongside the traditional asynchronous SAR logic.

When the prediction enable signal $V_{EN}$ is high, MUX2 pulls $\mathit{CLK}_{1-4}$ to VDD, causing the comparator clock to skip the first four comparison cycles. During this prediction phase, the four MSB capacitors are switched simultaneously according to the Kalman filter output. After the prediction phase completes, the SAR logic sequentially generates internal clock signals $\mathit{CLK}_{5-10}$ for the remaining six LSB conversions. When $V_{EN}$ is low, MUX2 selects the normal $\mathit{CLK}_{1-4}$ signals, and the circuit operates identically to a conventional asynchronous SAR ADC. This dual-mode design ensures a fail-safe fallback path: if the Kalman filter prediction is disabled or produces an out-of-range estimate, the ADC reverts to its full ten-cycle conventional conversion.

\subsection{DAC Control Logic}

\begin{figure}[!t]
\centering
\includegraphics[width=3in]{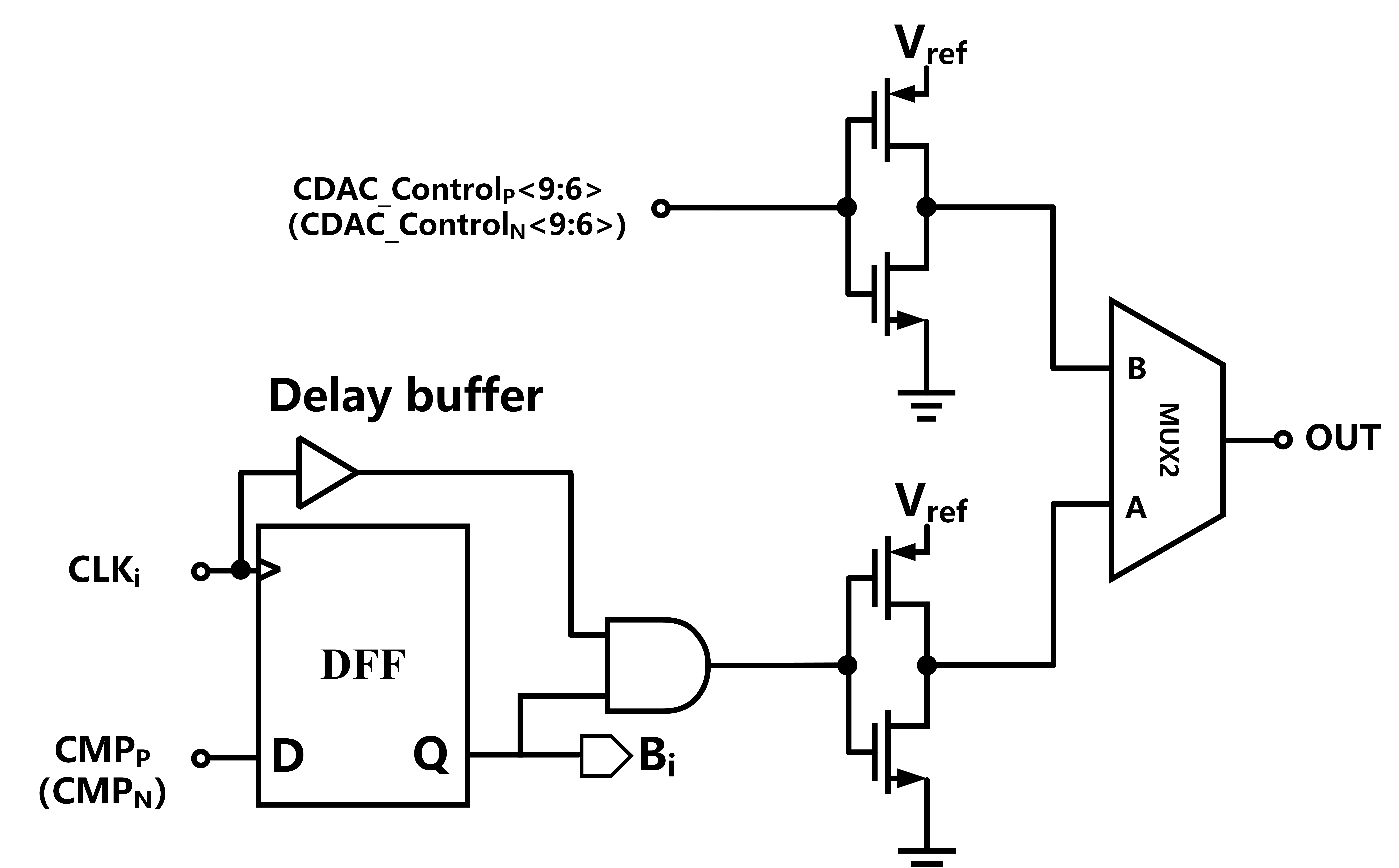}%
\caption{DAC switch control logic. Each MSB switch is driven by a MUX2 that selects between the comparator signal (traditional mode) and the Kalman filter prediction signal (predictive mode) based on $V_{EN}$.}
\label{fig6}
\end{figure}

Fig.~\ref{fig6} shows the capacitor switch control logic. In the proposed design, the 4-bit MSB capacitor switches can be driven by either the comparator outputs (traditional mode) or the Kalman filter prediction signals (predictive mode), selected through MUX2 multiplexers.

When $V_{EN}=1$, the Kalman filter is active and the comparator skips the four MSB comparisons. The filter signals directly control $\mathit{CDAC\_Control\_P}<9:6>$ and $\mathit{CDAC\_Control\_N}<9:6>$, driving the bottom plates of the positive and negative CDAC arrays, respectively. For instance, if $\mathit{CDAC\_Control\_P}<9>=1$, the corresponding switch pulls the bottom plate from $V_{\text{REF}}$ to ground; if $0$, it remains at $V_{\text{REF}}$. When $V_{EN}=0$, the MUX2 blocks route the comparator outputs to the switches, and the capacitor array flips bit by bit according to the conventional monotonic switching sequence.

% ============================================================
% SECTION V: Simulation Results
% ============================================================
% Section V: Simulation Results
\section{Simulation Results}
\label{sec:results}

The 10-bit ADC designed in this work was evaluated through behavioral simulations of the CDAC switching energy and through circuit-level simulations of the complete ADC core.

\subsection{Switching Energy}

\begin{figure}[!t]
\centering
\includegraphics[width=3in]{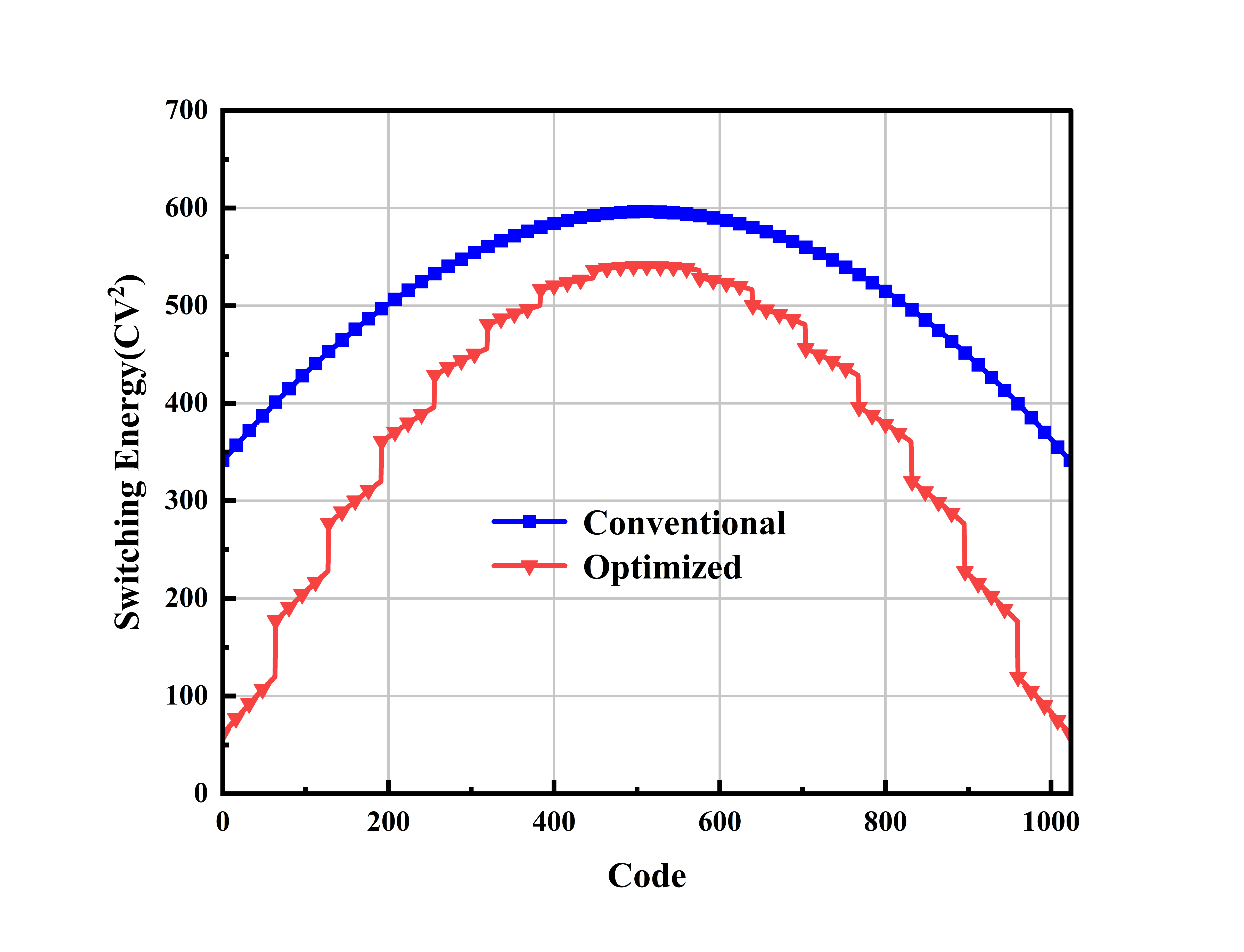}%
\caption{Simulated CDAC switching energy versus output code for a 10-bit ADC. The proposed method (predictive mode) is compared against the conventional monotonic switching scheme.}
\label{fig7}
\end{figure}

Fig.~\ref{fig7} compares the CDAC switching energy of the proposed method against the conventional monotonic switching scheme for a full-scale sinusoidal input. The proposed method reduced switching energy across the full output code range. The reduction reflects two design properties: (i) simultaneous MSB switching enabled by the Kalman filter prediction eliminates sequential charge-redistribution losses, and (ii) the optimized 4-bit switching logic minimizes the number of capacitor transitions per conversion.

\subsection{Power and Dynamic Performance}

\begin{figure}[!t]
\centering
\includegraphics[width=3in]{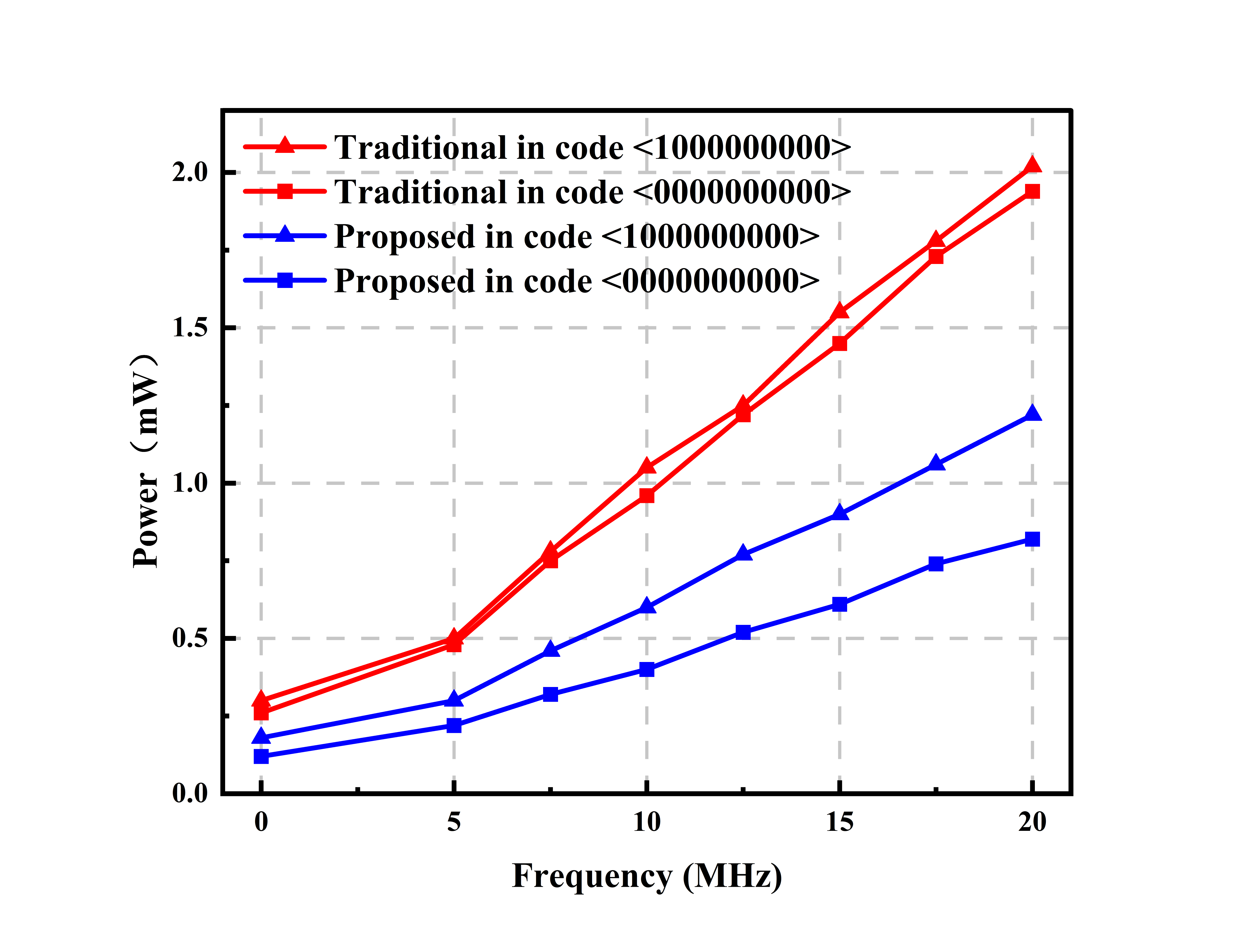}%
\caption{Measured total power consumption versus sampling frequency for the traditional mode (filter off) and predictive mode (filter on).}
\label{fig8}
\end{figure}

Fig.~\ref{fig8} shows the measured total power consumption of the ADC in both traditional and predictive modes as a function of sampling frequency. At 20~MS/s, the power consumption decreased from 1.96~mW (traditional mode) to 0.975~mW (predictive mode), a 50.3\% reduction.

\begin{figure}[!t]
\centering
\includegraphics[width=3in]{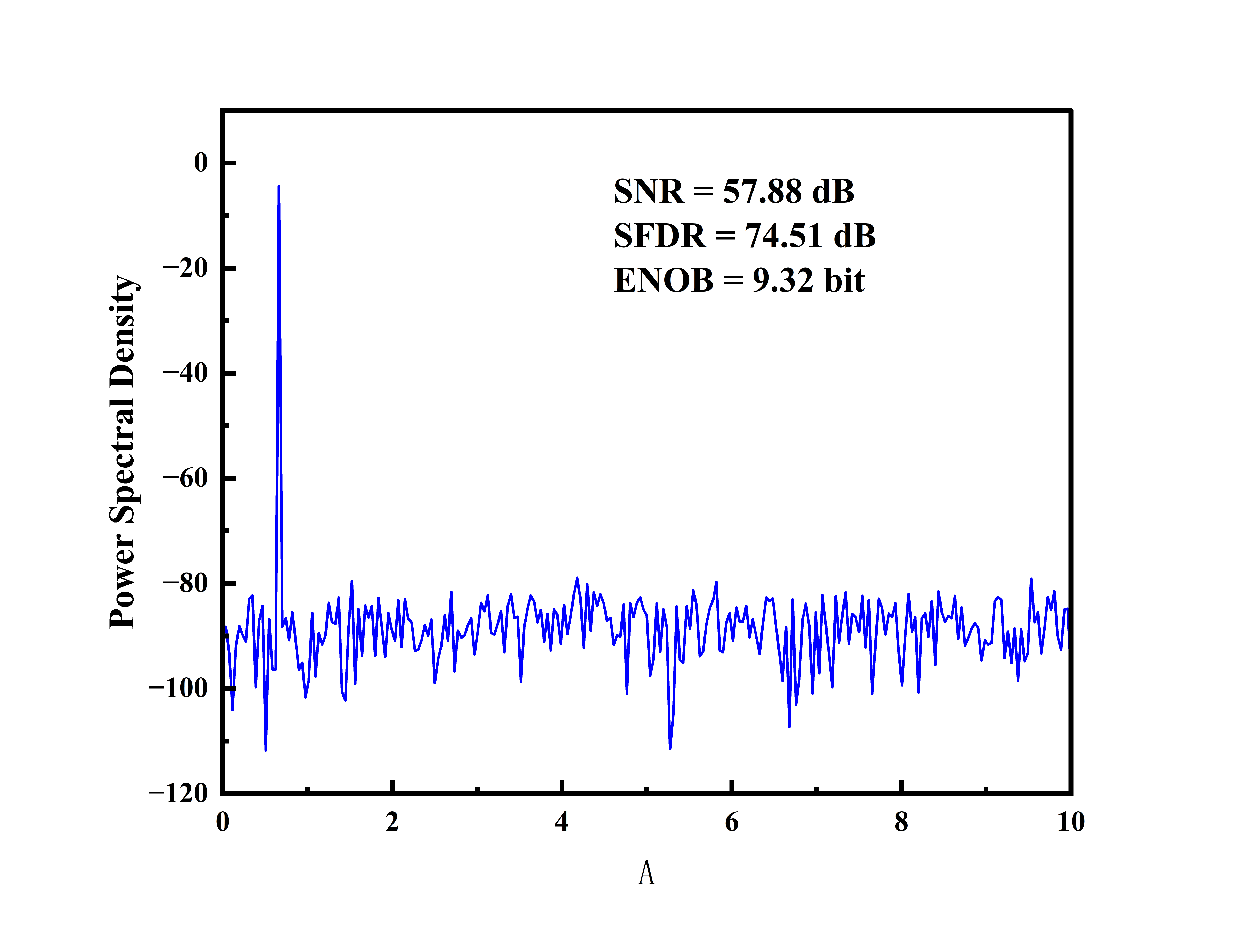}%
\caption{Output FFT spectrum of the proposed ADC sampling a 504~kHz input at 20~MS/s. The measured SNR and SFDR are 57.88~dB and 74.51~dB, respectively.}
\label{fig9}
\end{figure}

Fig.~\ref{fig9} presents the measured FFT spectrum with a 504~kHz input tone at a 20-MS/s sampling rate under a 1.8 V supply. The measured SNR and SFDR were 57.88~dB and 74.51~dB, respectively, consistent with the dynamic performance expected of a 10-bit ADC across the Nyquist band.

\subsection{Comparative Analysis of Predictive SAR ADC Architectures}

Table~\ref{T2} summarizes the performance of the proposed ADC alongside recently reported signal-specific SAR ADCs. The proposed design achieved the highest sampling rate among the compared works (20~MS/s) while attaining an FoM of 123.1~fJ/conv-step in predictive mode. The references selected for comparison target similar application spaces (biomedical and low-power sensing). The proposed work was the only design that offered a dual-mode configurable prediction scheme, capable of operating in both conventional and predictive modes.

\begin{table*}[t]
\centering
\caption{Performance comparison with state-of-the-art SAR ADCs.}
\label{T2}
\begin{tabular}{|c|c|c|c|c|}
\hline
 & TCASI~\cite{9} & JSSC~\cite{16} & ISCAS~\cite{17} & This Work \\
\hline
Technology & 90 nm & 0.18 $\mu$m & 0.18 $\mu$m & 0.18 $\mu$m \\
\hline
Supply Voltage (V) & 0.3 & 1.8 & 1.0 & 1.8 \\
\hline
Resolution (bit) & 10 & 10 & 8 & 10 \\
\hline
Sampling Rate (MS/s) & 0.15 & 0.2 & 0.25 & 20 \\
\hline
ENOB (bit) & 8.85 & 7.96 & 7.06 & 8.63 \\
\hline
Power & 67.3 nW   & 19 $\mu$W & 2.86 $\mu$W & 1.96 mW / 0.975 mW \\
\hline
FoM (fJ/conv-step) & 0.97 & 381 & 85.7 & 247.4 / 123.1 \\
\hline
\end{tabular}
\end{table*}

% ============================================================
% SECTION VI: Discussion and Future Work
% ============================================================
% Section VI: Discussion and Future Work
\section{Discussion and Future Work}
\label{sec:discussion}

\subsection{Hardware Implementation of the Kalman Filter}

In the present work, the Kalman filter was modeled at the algorithm level  to validate the feasibility of prediction-assisted MSB switching. A dedicated hardware implementation remains as future work. However, previous studies have demonstrated that low-order Kalman filters can be efficiently realized in digital hardware. For instance,~\cite{13} presents VLSI architectures that achieve 2.8$\times$ fewer arithmetic operations and 3.3$\times$ fewer clock cycles compared to prior art. Moreover,~\cite{14} reports an FPGA-based Kalman filter coprocessor achieving 7--12$\times$ acceleration relative to an ARM software implementation. Given that the proposed design requires only a second-order Kalman filter operating at the sampling rate, the additional digital area and power overhead are expected to be modest relative to the analog CDAC and comparator.

\subsection{Technology Scaling Benefits}

Scaling to advanced CMOS nodes would further improve the performance of this ADC architecture. Compared with the 180-nm process, nodes such as 40 nm or 28 nm offer lower transistor leakage, higher capacitance density, and reduced supply voltages (e.g., 0.6--0.9 V). Under these conditions, power consumption is expected to drop from the measured 0.975~mW (predictive mode at 180 nm) to the microwatt range, with an estimated 30--40\% improvement in energy efficiency.

\subsection{Limitations and Future Directions}

Several limitations of the current work merit discussion. First, the Kalman filter is currently verified only as a  behavioral model; comprehensive co-simulation with a synthesized RTL implementation of the filter and the ADC core is required to fully validate the system-level power and timing. Second, the prediction accuracy degrades during large-signal transients (e.g., ECG QRS complexes), where the residual may exceed the 64-LSB correction range. Under such conditions, the ADC relies on the dual-mode fallback mechanism described in Section~\ref{sec:control} to maintain conversion accuracy, albeit at the cost of temporarily increased power consumption. Third, the current design does not include an on-chip calibration mechanism for CDAC capacitor mismatch, which may limit the achievable linearity under process variations. Fourth, the comparison in Table~\ref{T2} spans a wide range of sampling rates; a more controlled comparison with designs at similar speeds (10--50 MS/s) would strengthen the benchmarking.

Future work will address these limitations through: (i) RTL implementation and synthesis of the Kalman predictor with post-layout co-simulation; (ii) an adaptive mode-switching controller with transient detection to minimize the duration of traditional-mode fallback; (iii) integration of foreground digital calibration for CDAC mismatch; and (iv) extension to implantable medical devices through on-chip integration with analog front-end preprocessing modules.

% ============================================================
% SECTION VII: Conclusion
% ============================================================
% Section VII: Conclusion
\section{Conclusion}

This paper has presented a Kalman filter-assisted data-predictive SAR ADC that reduces CDAC switching energy by predicting the first 4 MSBs before each conversion cycle. The proposed architecture enables simultaneous parallel switching of the MSB capacitors, eliminating redundant CDAC transitions and shortening the quantization cycle by 4 clock periods. An optimized 4-bit MSB switching scheme further suppresses residual redundant switching. Fabricated in a 180-nm CMOS process and configurable between traditional and predictive modes, the ADC achieves a 50.3\% power reduction at 20~MS/s under a 1.8-V supply when the predictive mode is engaged. Dynamic performance measurements confirm robust operation across the full Nyquist bandwidth, with an SNDR/SFDR of 53.63/65.89~dB at 504~kHz. The proposed predictive framework offers a promising low-energy data conversion solution for next-generation wireless sensor nodes and wearable health monitoring systems.

% ============================================================
% Bibliography
% ============================================================

\begin{IEEEbiography}{Michael Shell}
Use $\backslash${\tt{begin\{IEEEbiography\}}} and then for the 1st argument use $\backslash${\tt{includegraphics}} to declare and link the author photo. Use the author name as the 3rd argument followed by the biography text.
\end{IEEEbiography}

\vspace{11pt}
\bf{If you will not include a photo:}\vspace{-33pt}
\begin{IEEEbiographynophoto}{John Doe}
Use $\backslash${\tt{begin\{IEEEbiographynophoto\}}} and the author name as the argument followed by the biography text.
\end{IEEEbiographynophoto}

\vfill

\end{document}